\documentclass{ws-ijmpcs}

\usepackage{subfigure}
\newcommand{\wh}{\widehat}

\def\beq {\begin{equation}}
\def\eeq {\end{equation}}
\newcommand{\as}{\alpha_s}

\begin{document}

\markboth{D. Boito}
{Perturbative Expansion of $\tau$ Hadronic Spectral Function Moments}

\catchline{}{}{}{}{}
%
%%%%%%%%%%%%%%%%%%%%%%%%%%%%%%%%%%%%%%%%%%%%%%%%%%%%%%%%%%%%%%%%%%%%

\title{PERTURBATIVE EXPANSION OF $\tau$ HADRONIC SPECTRAL FUNCTION MOMENTS}

\author{DIOGO BOITO}

\address{Physik Department T31, Technische Universit\"at M\"unchen\\
James-Franck-Stra\ss e 1, D-85748 Garching, Germany\\
}

%diogo.boito@tum.de}

%\author{SECOND AUTHOR}

%\address{Group, Laboratory, Address\\
%City, State ZIP/Zone, Country\\
%second\_author@domain\_name}

\maketitle

%\begin{history}
%\received{Date}
%\revised{Day Month Year}
%\end{history}

\begin{abstract}
  In the extraction of $\alpha_s$ from hadronic $\tau$ decay data
  several moments of the spectral functions have been
  employed. Furthermore, different renormalization group improvement
  (RGI) frameworks have been advocated, leading to conflicting values
  of $\alpha_s$. Recently, we performed a systematic study of the
  perturbative behavior of these moments in the context of the two
  main-stream RGI frameworks: Fixed Order Perturbation Theory (FOPT)
  and Contour Improved Perturbation Theory (CIPT). The yet unknown
  higher order coefficients of the perturbative series were modelled
  using the available knowledge of the renormalon singularities of the
  QCD Adler function. We were able to show that within these RGI
  frameworks some of the commonly employed moments should be avoided
  due to their poor perturbative behavior. Furthermore, under
  reasonable assumptions about the higher order behavior of the
  perturbative series FOPT provides the preferred RGI
  framework.  
\keywords{$\tau$-decays, Renormalization Group, Renormalons}
\end{abstract}

\ccode{PACS numbers:13.35.Dx, 12.38.Cy}
\ccode{TUM-916/13}

\section{Introduction, framework, and results}

The determination of $\alpha_s$ from hadronic $\tau$ decays is one of
the most precise determinations of the QCD coupling.\cite{BNP92,DHZ05}
In the inclusive $\tau$ hadronic width, $R_\tau$, in spite of the
relatively low energy scale set by the $\tau$ mass, the
non-perturbative contribution is subleading and the theoretical
description is dominated by perturbative QCD.  In detailed $\alpha_s$
analyses, one exploits the knowledge of the spectral functions,
measured by OPAL and ALEPH at LEP,\cite{OPAL,ALEPH} in order to
construct additional observables. Different moments of the spectral
functions are used; their theoretical counterpart is evaluated
through  finite energy sum-rules, as    contour integrals in the
complex-energy plane. In this context, $R_\tau$ can be understood as
a particular choice of moment of the spectral functions integrated up to
the kinematical limit $s_0=m_\tau^2$. Other
analytic weight functions  and upper limits $s_0\leq m_{\tau}^2$
(as long as $s_0$ is large enough to allow a perturbative treatment)
also define  observables that can be computed theoretically. The use of
tailored weight functions can be instrumental to the $\alpha_s$
analysis, e.g., suppressing or enhancing the non-perturbative
contributions.\cite{OPAL,ALEPH,ALEPH2008,MY2008,AlphasDVs2011,AlphasDVs2012}

Our  focus is on the perturbative QCD contribution to the different
moments used in $\alpha_s$ analyses. Recently,\cite{BBJ2012,Boito2013} we investigated the
convergence of the perturbative series after integration in the
complex plane employing two different renormalization group improvement
(RGI) prescriptions and discussed how the convergence properties of the
series depend on the specific moment used. Here we briefly describe the methods
and the results of our analysis and try to summarize the main
conclusions.

For the theoretical description, the relevant quantity is the QCD Adler function,
which is renormalization group (RG) invariant.  The perturbative
contribution to the observable defined by the weight function
$w_i(s)$, denoted $\delta^{(0)}_{w_i}$, is obtained through an
integration  on the complex energy
plane along the circle of radius $s_0$.  Defining $x=s/s_0$,
$W_i(x)=2\int_x^1\, w_i(z)dz$, and $a_\mu\equiv a(\mu^2)\equiv\alpha_s(\mu)/\pi$, the explicit expression reads
\begin{equation}
\label{del0}
\delta^{(0)}_{w_i} \,=\, \sum\limits_{n=1}^\infty  \sum\limits_{k=1}^{n}
k\,c_{n,k} \;\frac{1}{2\pi i}\!\!\oint\limits_{|x|=1} \!\! \frac{dx}{x}\,
W_i(x) \log^{k-1}\biggl(\frac{-s_0 x}{\mu^2}\biggr)a_\mu^n .
\end{equation}
The dynamical input to this series is fully contained in the $c_{n,1}$
coefficients, known at present up to $\alpha_s^4$
order.\cite{BCK08} The other coefficients can be determined using
RG invariance in terms of the $c_{n,1}$ and $\beta$-function
coefficients.

The scale $\mu$ in
the last equation can  be set in a convenient way due to RG invariance. The two mainstream choices are known as fixed-order perturbation theory\cite{Jamin05} (FOPT) obtained by fixing the scale
$\mu^2=s_0$, and contour improved perturbation theory\cite{CIPT,PLD92}
(CIPT) obtained when the running of $\alpha_s$ is resummed along the
contour by setting $\mu^2 = x s_0$.  (The FOPT series can be reobtained from CIPT
via  the  expansion of  the running coupling $a(x s_0)$ in terms of
the coupling at a fixed scale $\mu^2 =s_0$.)
Both expansions are expected to
diverge at large orders due to factorial growth of the perturbative
coefficients. Therefore, the two series define two different
asymptotic expansions (at best) to the  value of the
$\delta^{(0)}_{w_i} $. In practice, the numerical differences are
large at $\alpha_s^4$ which represents one of the dominant sources of
theoretical uncertainty.

A comparison between the two approaches regarding their success in
approximating $\delta^{(0)}_{w_i}$ depends on assumptions about the higher
order terms.  A strategy to deal with this problem based on the
available knowledge of the renormalon singularities of the Borel
transformed Adler function was put forward by Beneke and
Jamin.\cite{BJ2008} They were able to show that under reasonable
assumptions --- to be discussed below --- FOPT is to be preferred for
the inclusive  $\tau$ hadronic width. Later we extended this
analysis\cite{BBJ2012,Boito2013} in order to ascertain how the
behavior of the perturbative series depends on the moment $w_i(x)$
as well as on the value of $s_0$. 

We work with the Adler function $\wh D$, which contains only the corrections to the  parton model result, and define its Borel transform $B[\wh D](t)$ as
\beq
 \wh D(s) \,\equiv\, \sum\limits_{n=0}^\infty r_n \,\as(\sqrt{s})^{n+1} \,, \qquad \mbox{and}\qquad B[\wh D](t) \,\equiv\, \sum\limits_{n=0}^\infty r_n\,\frac{t^n}{n!},
\eeq
with $c_{n,1} = \pi^nr_{n-1}$. The original series $\wh D$ can be
understood as an asymptotic expansion of the inverse of $B[\wh D](t)$
\beq
\wh D(\alpha) \equiv \int_0^\infty e^{-t/\alpha}B[\wh D](t)\label{BorelSum},
\eeq
when the integral exists.   This equation defines the Borel sum of the series.

The Borel transformed Adler function has singularities along the real
axis both for negative and positive values of $t$, known as renormalon
singularities (for a review see\cite{Renormalons}). General RG
arguments and the structure of the OPE allows one to determine the
position and, in principle, the strength of these singularities --- the
residues are unknown.  Singularities on the positive real axis,
infrared (IR) renormalons, give rise to fixed sign divergent
series. These singularities obstruct the integration in Eq.~(3) and
produce an ambiguity in the Borel sum related to the prescription used to 
define the integral.  Singularities for $t<0$, ultraviolet (UV) renormalons,
give rise to sign alternating divergent series. The fixed sign nature
of the exactly known coefficients of the Adler function suggests that
the series is dominated by IR singularities at low and intermediate
orders.

The strategy then consists in constructing a model for the Borel
transformed Adler function containing a small number of dominant
renormalon singularities whose residues are unknown. The residues are
then fixed in order to reproduce the known coefficients and an estimate of
the $\alpha_s^5$ term.  The Adler function can  be reconstructed to
all orders and the RG improved result can be compared with a ``true"
result for $\delta^{(0)}_{w_i}$, obtained using Eq.~(\ref{BorelSum}).
The main assumptions behind this strategy is that the series exhibits
some regularity, and that sufficiently many terms are known in order
to fix the contribution of the leading renormalons.  This has been
tested in detail using the large-$\beta_0$ limit of QCD and the
plausibility of these assumptions has been confirmed.\cite{BBJ2012}
\begin{figure}[!t]
\begin{center}
%\subfigure[$w(x)=1-x^2$, FOPT.]{\includegraphics[width=.49\columnwidth,angle=0]{./Figs/Alls0sNorPertSeries_CM_FOPT_W1mxsq}\label{RMW1mxsqFO}}
%\subfigure[$w(x)=1-x^2$, CIPT.]{\includegraphics[width=.49\columnwidth,angle=0]{./Figs/Alls0sNorPertSeries_CM_CIPT_W1mxsq}\label{RMW1mxsqCI}}
\subfigure[$w_\tau(x)=(1-x)^2(1+2x)$, FOPT.]{\includegraphics[width=.49\columnwidth,angle=0]{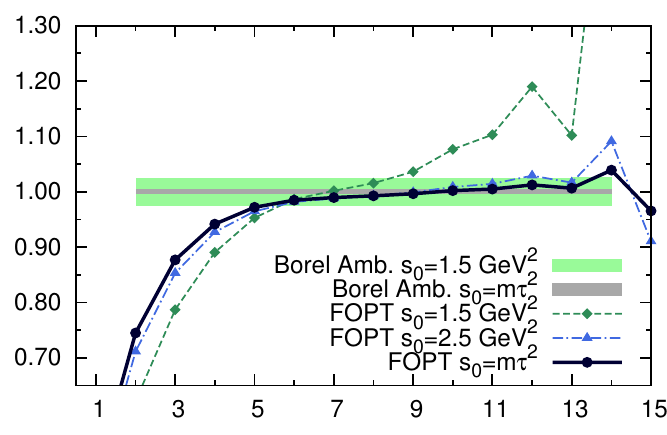}\label{RMWtauFO}}
\subfigure[$w_\tau(x)=(1-x)^2(1+2x)$, CIPT.]{\includegraphics[width=.49\columnwidth,angle=0]{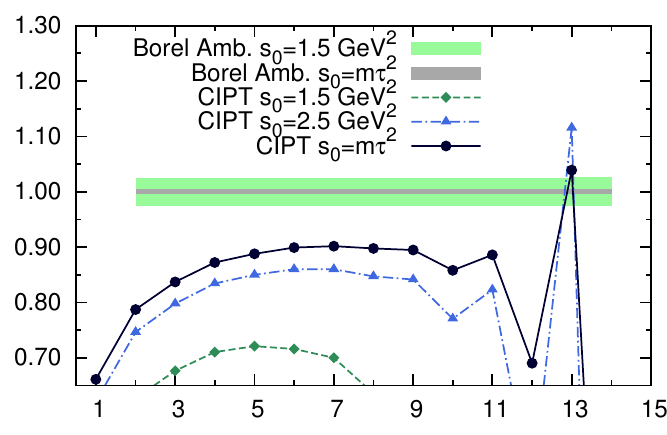}\label{RMWtauCI}}
\caption{Reference model. $\delta^{(0)}_{w_\tau}(s_0)$ order by order in $\alpha_s$ normalised to the Borel sum  for FOPT (left)  and CIPT (right) with three values of $s_0$:  1.5~GeV$^ 2$, 2.5~GeV$^ 
2$, and $m_\tau^ 2$. Bands give the Borel ambiguities. }\vspace{-0.8cm}
\label{fig:IdealMom}
\end{center}     
\end{figure}

\begin{figure}[!ht]
\begin{center}
%\subfigure[$w(x)=1-x^2$, FOPT.]{\includegraphics[width=.49\columnwidth,angle=0]{./Figs/Alls0sNorPertSeries_AM_FOPT_W1mxsq}\label{AMW1mxsqFO}}
%\subfigure[$w(x)=1-x^2$, CIPT.]{\includegraphics[width=.49\columnwidth,angle=0]{./Figs/Alls0sNorPertSeries_AM_CIPT_W1mxsq}\label{AMW1mxsqCI}}
\subfigure[$w_\tau(x)=(1-x)^2(1+2x)$, FOPT.]{\includegraphics[width=.49\columnwidth,angle=0]{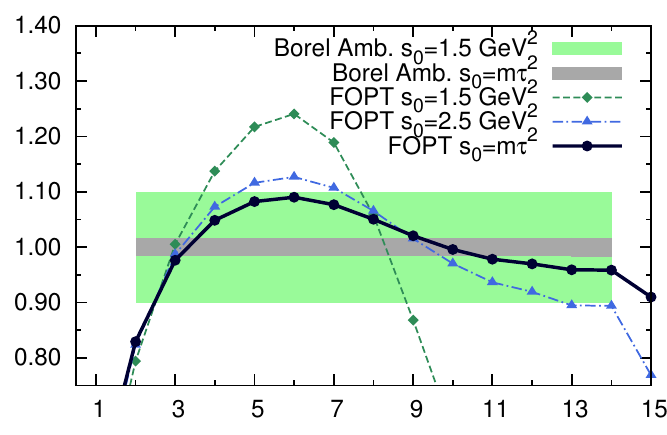}\label{AMWtauFO}}
\subfigure[$w_\tau(x)=(1-x)^2(1+2x)$, CIPT.]{\includegraphics[width=.49\columnwidth,angle=0]{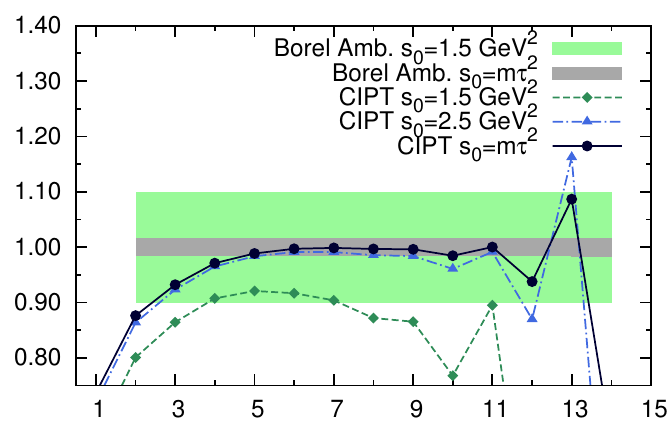}\label{AMWtauCI}}
\caption{Alternative model. $\delta^{(0)}_{w_{\tau}}(s_0)$ order by order in $\alpha_s$ normalised to the Borel sum  for FOPT (left)  and CIPT (right) with three values of $s_0$:  1.5~GeV$^ 2$, 2.5~GeV$^ 
2$, and $m_\tau^ 2$. Bands give the Borel ambiguities.}\vspace{-0.5cm}
\label{fig:IdealMomAM}
\end{center}     
\end{figure}

Two models were constructed.\cite{BBJ2012,BJ2008} The first, and more
realistic one in our opinion, assumes a logical hierarchy between the
IR renormalon contributions. It includes the two leading IR
renormalons along with the leading UV (whose signature is not yet seen
in the known coefficients).  This gives rise to a model that we dub the
{\it reference model} (RM). This model is contrasted with  a second
where the leading IR renormalon is artificially removed from the Borel
transformed, which realizes the case where the contributions of the
second IR renormalon are maximized. This is called the {\it
  alternative model} (AM). Using these two models for the higher
orders, we performed a systematic analysis of a collection of
different moments, using different $s_0$ values, and comparing the
performance of FOPT and CIPT. As an example, Figs.~1 and~2 show
results for the kinematic moment within the two
models.\cite{BBJ2012,Boito2013} They clearly show the preference for
FOPT within the more realistic RM.  CIPT gives the better
approximation when the leading IR renormalon is artificially
suppressed.

\section{Conclusions}

Here we try to summarize our main conclusions.\cite{BBJ2012,Boito2013,BJ2008}

\begin{itemlist}

\item The finiteness of the radius of convergence of the expansion of
  the running coupling\cite{PLD92} in terms of $\alpha_s(\sqrt{s_0})$
  does not disfavor FOPT, for the perturbative series of $\wh D$
  is expected to be divergent. The models corroborate this
  conclusion:  FOPT provides a better asymptotic expansion than
  CIPT in the case of the RM.

\item CIPT and FOPT define two different asymptotic series.  FOPT
  treats the running of $\alpha_s$ and the $c_{n,1}$ coefficients on
  an equal footing and only includes at a given order $n$ terms up to
  order $\alpha_s(\sqrt{s_0})^n$. In CIPT  the running of $\alpha_s$
  is always resummed to all orders although only a finite number of
  $c_{n,1}$ coefficients contribute at a given order.  Contrary to
  what is often stated, there is no reason to believe that the
  differences in the $\alpha_s$ values from FOPT and CIPT can be
  attributed to missing higher orders.

\item The preference for FOPT or CIPT can be mapped into an assumption
  about the renormalon content of the QCD Adler function. FOPT is
  superior whenever a sizable contribution from the leading ($D=4$) IR
  renormalon is present.  The naturalness of this scenario is a strong
  argument in favor of FOPT. The (artificial) suppression of this
  contribution realizes a scenario where CIPT
  is superior.

\item In the context of the RGI frameworks discussed here, some of the
  moments that are commonly employed in determinations of $\alpha_s$
  should be avoided due to their poor perturbative behavior. In
  particular, the moments that emphasise higher OPE condensates
  ($D\geq 8$) used  in\cite{OPAL,ALEPH,ALEPH2008} should be avoided.
  In contrast, the moments used
  in\cite{MY2008,AlphasDVs2011,AlphasDVs2012} have a better
  convergence, and at least one of the series (FOPT or CIPT)
  approaches the ``true" value at relatively
  low orders.

\item Ideally, the goodness of the RGI framework should be moment
  independent. Preferentially, it should also be independent on the
  assumptions about the renormalon singularities of the Adler function (in
  the context of our work this can be rephrased as being {\it model
    independent}). To our knowledge, the most promising
  strategy in this direction is the use of conformal mapping techniques based on the
  partial knowledge of the Borel transformed Adler function.\cite{Caprinietal} 

\end{itemlist}

\vspace{-0.5cm}
\section*{Acknowledgements}

It is a pleasure to thank Martin~Beneke and Matthias~Jamin for the
collaboration in this subject and for the reading of the manuscript.
This work was supported by the Alexander von Humboldt Foundation.

%\begin{thebibliography}{000} %for 3 digits
%\begin{thebibliography}{00}  %for 2 digits

\end{document}